# Visualization of Ion Channels in Membranes using Electrochemical Strain Microscopy

*Suran Kim,*[‡a] *Chungik Oh,*[‡a] *Hongjun Kim,*[‡a] Yuanyuan Shi,[b] Mario Lanza,*[*c] Kwangsoo No*[a], and Seungbum Hong*[a,d]

[a]*Department of Material Science and Engineering, KAIST, Daejeon 34141, Korea*
[b]*Materials Science and Engineering Department, Guangdong Technion – Israel Institute of Technology, Shantou, China*
[c]*Institute of Functional Nano and Soft Materials, Soochow University, Suzhou, China*
[d]*KAIST Institute for NanoCentury, KAIST, Daejeon 34141, Korea*

[*]Corresponding authors: mario.lanza@gtiit.edu.cn, ksno@kaist.ac.kr, seungbum@kaist.ac.kr

**Abstract**: Polymer composite electrolytes of Nafion and phosphotungstic acid (PWA) are fabricated and analyzed using electrochemical strain microscopy (ESM) and conductive atomic force microscopy (C-AFM) to visualize hydrophilic ion channels near the surface, which are composed of water and sulfonic acid groups. The results indicate that the fibrillar objects in ESM image, without significant changes in topography, are hydrophilic ion channels and additional ion channels formed by interaction between PWA and sulfonic groups in Nafion. In this study, the buried ion channels lying under the surface are probed as well as the inlet and outlet of the channels on the surface through combined use of ESM and C-AFM. The results further enhance the understanding of ionic conduction in composite polymer electrolytes in various fields.

Nafion membranes, due to high proton conductivity, selective permeability to water, and excellent chemical stability, have drawn considerable attention as a polymer electrolyte in various fields from energy applications to bio-inspired and biomedical applications.[1–11] Nafion membranes show a bi-continuous nanostructure containing a hydrophobic matrix and hydrophilic water-ion channels that allow transport of ions and water.[3,6,12–15] The nanostructure within the membranes is of significant research interest because it is intimately related to the ion transport, which consequently affects the functionality for various applications from fuel cells to artificial muscles.[13]

HPA (Heteropoly acid) is a class of acid made up of a particular combination of hydrogen and oxygen with certain metals and non-metals and a subset of well-known inorganic metal oxides. The most thoroughly studied HPAs is $H_3PW_{12}O_3$ (PWA; phosphotungstic acid), which consists of a centrally located, four-coordinate phosphorus linked via oxygen atoms to 12 tungsten-oxygen.[16] Due to high intrinsic proton conductivity and good water retention, PWA is widely used in hybrid organic-inorganic composite electrolytes for fuel cell application.[17–21] The addition of PWA redistributes the hydrophilic channels to a more interconnected morphology due to the interfacial interaction between PWA and side chain of Nafion. PWA insertion in Nafion membranes increases the number of active sites in the membrane, as well as the number of water molecules by the addition of sulfonic groups, which results in the improvement of proton transport through the membrane and the electrochemical properties of the membrane.[18,22–24]

There have been many research reports on the nanostructures within the membranes using experimental methods such as X-ray scattering and computer simulation tools.[3,14,25–27] In these studies, researchers concluded that the long-range nano-channels composed of aggregated and hydrated ionic groups are embedded in a continuous matrix.[25,28] Atomic force microscopy (AFM) is a promising strategy for imaging the morphology, mechanical and electrochemical properties at the nanoscale. Especially, conductive AFM has been used to probe local conductivity and the distribution and density of ionic clusters.[29–33] By applying bias voltage across the membrane using a conductive AFM tip, the local currents produced by ionic and electronic currents inside the membranes were measured. However, it is a challenging task to obtain only ionic transport component from the electrical current signal.

Alternative solution is to use electrochemical strain microscopy (ESM) that can measure the bias-induced strain change on a surface using a conductive AFM tip. ESM has been used extensively to study lithium ion diffusion in lithium-conducting ceramics for battery applications.[34–39] Recent investigation demonstrated that the proton distribution and relaxation behavior in Nafion could be visualized and the local strain caused only by ions could be measured using ESM techniques. [40] Herein, we employ ESM as a tool for imaging simultaneously the topography and the local strain change caused by the electric field. By imaging the local strain, we could visualize the ion channels buried beneath the surface, thereby visualizing the nanostructure of Nafion/PWA composite membrane.

**Figure 1** shows the simultaneously obtained topography, ESM amplitude and phase images of composite membrane of phosphotungstic acid (PWA) and Nafion. At a glance, the ESM amplitude and phase show similar features when compared with the topography. However, a careful observation reveals that fibrillar objects with various lengths are found where there are no significant changes in topography. We speculate that these objects are hydrophilic ion channels because their shape and size are similar to thin string-like channels in the free bulk-like water distribution measured by atomic force microscope infrared-spectroscopy (AFM-IR) chemical imaging [41]. The discrepancies between channel sizes measured by AFM-IR, small-angle X-ray scattering (SAXS) [3] and ESM could be due to the different spatial resolutions of the characterization methods.

The fibrillar objects clearly show higher amplitude and phase values than those of matrix as shown in **Figure 1**(d) and (e), which suggests that more water and protons exist near the fibrillar objects. [40] There are three major peaks in ESM phase (see blue arrows in **Figure 1**(e)), which show no significant topography changes. In addition, three amplitude peaks (see red arrows in **Figure 1**(d)) were observed at the same location of the phase peaks. Our results support the claim that these objects are hydrophilic ion channels, which are composed of water and sulfonic acid groups in Nafion. No topography changes in these regions indicate that these ion channels are buried below the sample surface. Consequently, we observed the buried hydrophilic ion channels using ESM.

To confirm that the fibrillar objects in ESM images are hydrophilic ion channels, the morphology and current images of Nafion/PWA (3 wt%) composite were simultaneously obtained as shown in **Figure 2**(a) and (b). The current distribution in **Figure 2**(b) is spatially inhomogeneous. The current is associated with an electrochemical reaction at the tip and proton flow through the membrane as well as the transport of either electrons or holes. First of all, the current image shows similar features when compared with the topography. To figure out the relationship between the current and topography, the line profile of height and current at the same position is presented in **Figure 2**(c). There is no significant relevance between topography and current.

The areas with high (> 120 nA) current are presented in **Figure 2**(d), where no fibrillar objects were found. Instead, we could find circular bright spots in the current images as shown in **Figure 2**(d). The bright spots show the places where the AFM tip directly contacts the ionic networks. As the C-AFM detects only the currents that flow out of the surface, the bright spots indicate either the inlet or outlet of the ionic channels if the current comes solely from the proton transport. It should be noted, however, that the bright spots in the current images could be due to the electron or hole flow from e.g. impurities at the sample surface. In conductive

AFM, it is very difficult to distinguish the ionic and electronic contribution to the current signal, whereas ESM identifies the motion of ions that contribute to the local strain (see **Table 1**).

To explain these results, we present a schematic of contrast mechanisms responsible for the images we obtained in **Figure 3** (a) and (b). Nafion consists of hydrophilic water-ion channels and crystallites in non-crystalline matrix. We simplify the 3D structure near the membrane surface as schematically illustrated in **Figure 3** (c). ESM and C-AFM image could be obtained by scanning AFM tip with bias voltage. Here, we assumed that AFM tips with bias voltage scanned the line (d) that shows the cross-sectional image of our membrane containing buried ion channel near the surface. **Figure 3** (e) shows the expected ESM signal, contributed to by the local strain induced by sub-surface ion. The current signal will be detected by only direct contact to ionic channels at the surface. Therefore, the expected current signals will show only inlet and outlet of the channel, which are depicted in **Figure 3** (f). Although AFM tips scan the same area, ESM can probe the whole fibrillar ion channels lying under the surface, whereas C-AFM detects the inlet of outlet of the channels on the surface due to different sources of signal. Using ESM and C-AFM altogether, we could visualize the fibrillar ion channels near Nafion surface as well as those exposed to surface, leading to a complete map of ionic channels near the surface.

In summary, we demonstrated that ESM is an effective technique for visualizing hydrophilic ion channels near the surface of membranes. Using ESM, we could visualize the fibrillar ion channels near the surface, which are composed of water and sulfonic acid groups in Nafion. Besides these channels, the other ion channels might be formed by the interaction between phosphotungstic acid and sulfonic groups in Nafion. Overall, these results show that combined use of ESM and C-AFM can be a powerful in-situ tool for probing spatially distributed nanoscale ionic channels and further provide insight into improving performance of these materials in various applications, including fuel cells and artificial muscles.

**Experimental Section**

We used commercially available 10% Nafion solution (Dupont DE-1021) and phosphotungstic acid hydrate (PWA; $H_3PW_{12}O_{40} \cdot xH_2O$) – (Sigma-Aldrich). Membranes were prepared by recast procedure. The solvents of Nafion solution were exchanged by high boiling point solvent (DMF; Dimethylformamide). PWA (1, 2, 3, 4, 5 wt% based on solid Nafion) was dispersed in DMF by stirring to avoid particle aggregation. We added the PWA dispersed solution to Nafion solution and mixed them for 24 hours. Then the mixed solution was poured onto a casting mold. The solution in the mold was dried at 80 °C for 12 hours and cured at 120 °C for 6 hours. Fabricated membranes in the mold were peeled off from the mold surface. To remove their organic and inorganic impurities on the surface, the membranes were immersed in 2 N HCl (HCl : Hydrochloric acid) for 30 min at 80 °C. We immersed the membranes in water for 30 min at 80 °C to clean and hydrate them. The hydrated membranes were dried at 70 °C for 12 hours in vacuum oven.

ESM measurement was performed under ambient conditions using a commercial atomic force microscopy; AFM (XE-100, Park Systems) equipped with a lock-in amplifier (SR830, Stanford Research System) using Pt/Ir coated cantilevers (ATEC-CONTPt tip, NANOSENSORS). To enhance the sensitivity of the measurement, ESM measurements were conducted at contact resonance frequency. The high frequency ac voltage (0.7 $V_{pk}$ at 53 ~ 56 KHz) was applied to the samples to measure their electromechanical response. A typical scan rate used for the measurements was 0.3 Hz. Current imaging under sample bias voltage of -2 V was also performed using commercial AFM (XE-100) using Pt/Ir coated cantilevers with scan rate of 0.5 Hz.

**Acknowledgements**

This work is partly supported by Basic Science Research Programs (No. 2015R1D1A1A01056983) through the NRF, Korea, funded by the Ministry of Education, (No.

2018R1A2B6002194) through the NRF, Korea funded by the Ministry of Science and ICT, Big Sciences Research Program (No. 2017M1A2A2044498) through the NRF, Korea funded by the Ministry of Science and ICT and International Collaborative Energy Technology R&D Program (No. 20158510050020) through the KETEP, Korea funded by the Ministry of Trade, Industry and Energy.

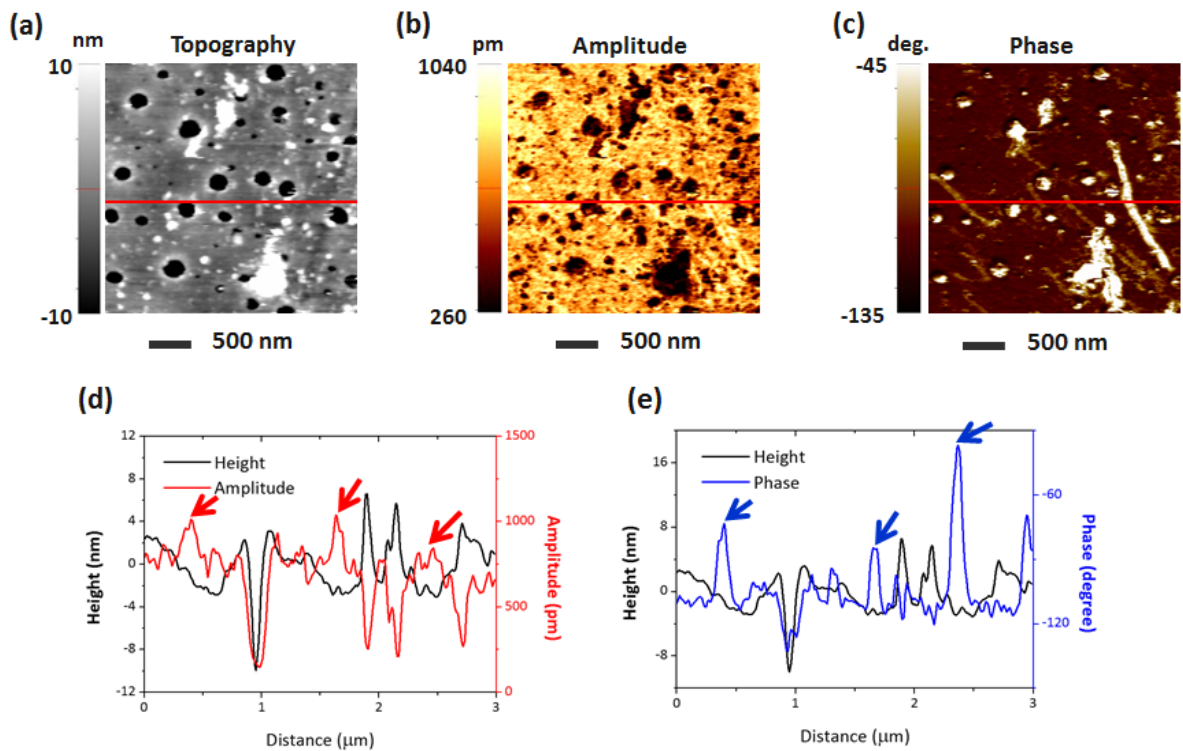

**Figure 1**. (a) Topography, (b) ESM amplitude and (c) ESM phase images of composite membranes, (d) line profile of height (topography) and ESM amplitude and (e) line profile of height (topography) and ESM phase (red line in (a), (b) and (c)). The red and blue arrows indicate the peaks in ESM signal, which show no significant topography changes.

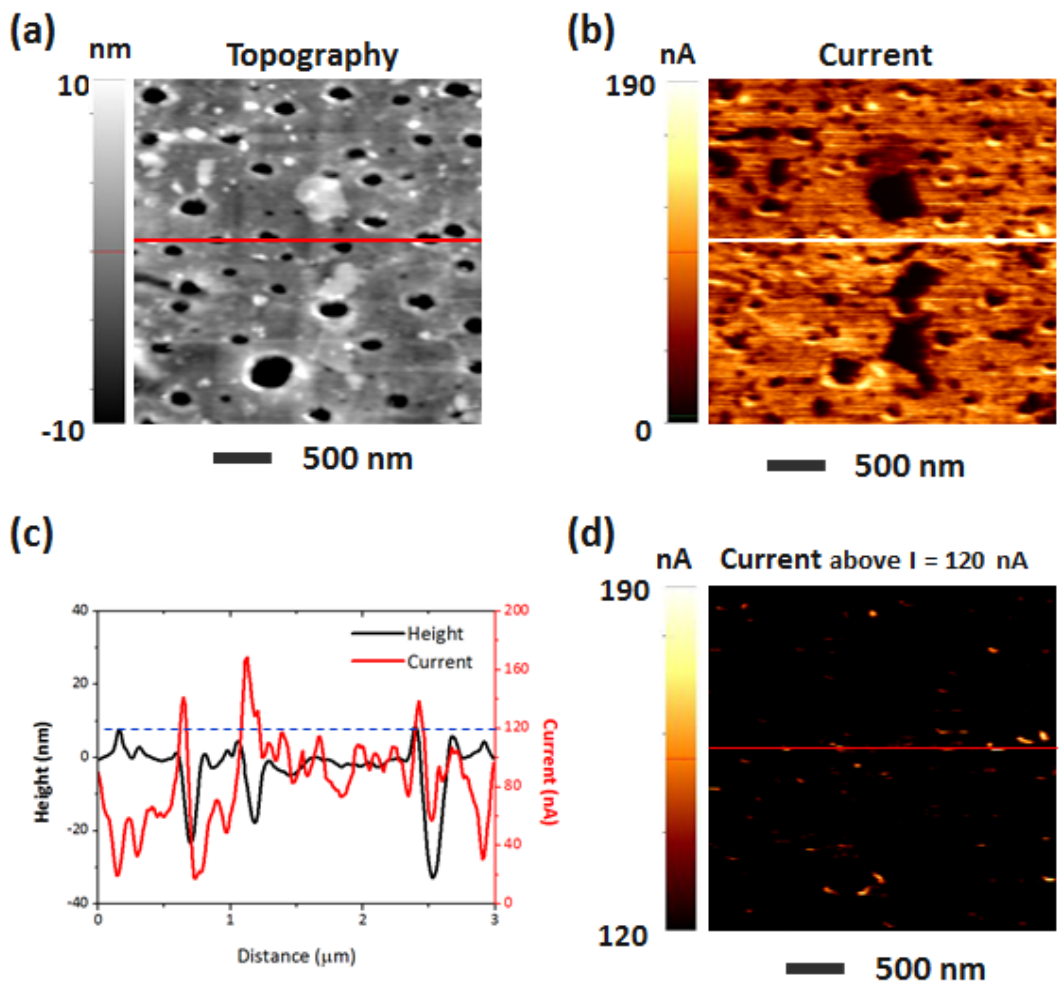

**Figure 2**. (a) Topography and (b) current images of composite membrane. (c) Line profiles of height (topography; red line in (a) image) and current images (white line in (b) image) and (d) current images with current above I = 120 nA.

**Table 1.** Comparison of C-AFM and ESM

|  | **Conductive AFM** | **ESM** |
|---|---|---|
| **Sources of contrast** | Electrons, Holes and Ions | Ions |
| **Location sensitivity** | Surface | Sub-surface |

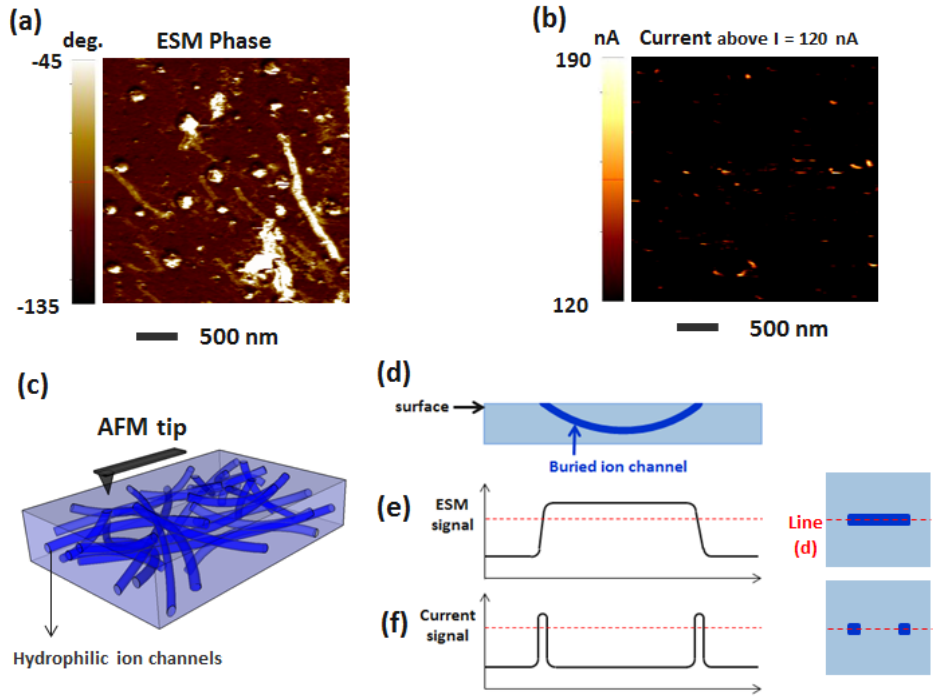

**Figure 3**. (a) ESM phase and (b) current images of membrane. Schematic illustration of (c) 3D structure near the membrane surface. The cross-sectional schematics (d) containing scanned line (d). ESM signal (e) and current signal (f) profile of line (d). ESM images (e) and current images (f) containing line (d).